\documentclass[aps,prl,showpacs,twocolumn,floatfix,superscriptaddress]{revtex4-1}
\usepackage{graphicx}
\usepackage{amsmath}
\usepackage{epstopdf}
\usepackage{mathbbol}
\DeclareMathAlphabet\mathbfcal{OMS}{cmsy}{b}{n}
\usepackage{color}

\begin{document}


\title{Hydrodynamic Equations for  Flocking Models without Velocity Alignment}

\date{\today}

\author{Fernando Peruani}
\affiliation{Universit{\'e} C{\^o}te d'Azur, Laboratoire J.A. Dieudonn{\'e}, UMR 7351  CNRS, Parc Valrose, F-06108 Nice Cedex 02, France}

\begin{abstract}
%
The spontaneous emergence of collective motion patterns is usually associated with the presence of a velocity alignment mechanism that mediates the interactions among the moving individuals. 
Despite of this widespread view, it has been shown recently that several flocking behaviors can  emerge in the absence of velocity alignment and as a result of short-range, position-based, attractive forces 
that act inside a vision cone. 
Here, we derive the corresponding hydrodynamic equations of a microscopic position-based flocking model, reviewing and extending 
previously reported results. 
In particular, we show that three distinct macroscopic collective behaviors can be observed:  
i) the coarsening of aggregates with no orientational order, ii) the emergence of static, elongated nematic bands, and iii) the formation of moving, locally polar structures, which we call worms.
The derived hydrodynamic equations indicate that active particles interacting via position-based interactions belong to a distinct class of active systems    
fundamentally different from other active systems, including velocity-alignment-based flocking systems. 
\end{abstract}



\maketitle


\section{Introduction}

The emergence of self-organized patterns of actively moving entities, from bacteria to sheep~\cite{vicsek2012,marchetti2013,yates2009,ballerini2008,gautrais2012,ginelli2015,toulet2015} 
and including human-made active systems~\cite{grossman2008,deseigne2010,weber2013,dauchot2015},  
are systematically explained invoking the presence of  some velocity alignment mechanism that mediates  the interactions among the moving individuals.   
This widespread view on collective motion patterns finds its roots in the  so-called Vicsek-like models~\cite{vicsek1995}  
extensively used to study flocking patterns~\cite{vicsek2012,marchetti2013}. 
The popularity of these models may be related to the fact that they   
 represent a very appealing playground for theoretical physicists given 
the Vicsek model's direct connection to one of the cornerstone models of equilibrium statistical physics: the XY model~\cite{doi}.     
While some nonequilibrium extensions of the XY model, including the diffusive XY spin model~\cite{peruani2010_xy,grossmann2016a},    
are susceptible of being mapped to their equilibrium counterpart,  
flocking models with velocity alignment, such as the original time-discrete Vicsek model~\cite{vicsek1995}
 and its continuum time version~\cite{peruani2008}, are fundamentally different. 
In idealized homogeneous media, these systems exhibit  long-range orientational order in two dimensions~\cite{vicsek1995,toner1995,toner1998} and 
the presence of anomalous  density fluctuations~\cite{ramaswamy2003,ramaswamy2010}. 
Although it has been recently shown that  the introduction of a few spatial heterogeneities restores a seemingly equilibrium-like behavior with quasi-long-range order  and normal fluctuations in two dimensions~\cite{chepizhko2013,chepizhko2015},  important differences (in two dimensions) remain in both homogeneous and heterogeneous media: the convective transport dictated by the orientation of the spin seems to prevent the emergence of topological defects. 

\begin{figure}
 \begin{center}
 \resizebox{9cm}{!}{\rotatebox{0}{\includegraphics{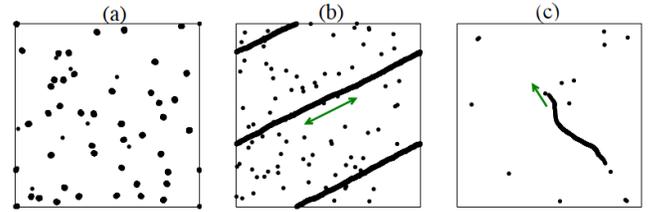}}}
 \caption{(Color online) Active particles interacting by short-range, attractive force acting inside a vision cone 
 self-organize into three distinct macroscopic patterns: (a) aggregates with no orientational order [$\beta=2.5$, $\sqrt{2D_\theta}=0.12$], 
 (b) nematic bands [$\beta=1.85$, $\sqrt{2D_\theta}=0.84$], and (c) moving, locally polar, structures called worms [$\beta=1.0$, $\sqrt{2D_\theta}=0.12$].  
 The double arrow in (b) indicates that inside a nematic band particles move in both directions.  
 The single arrow in (c) indicates the moving direction of the ``head" of the worm. 
 Panels (a), (b) and (c) corresponds to simulations snapshots of the model defined by Eqs.~(\ref{eq:mov_2d}) with $N=10000$ particles in a box of linear size $L=100$ with periodic boundary conditions with $v_0=1$ and $\gamma=5$. 
 }
 \label{fig:patterns}
 \end{center}
 \end{figure}


While the relevance of flocking models based on velocity alignment is undisputed in the realm of active matter and nonequilibrium statistical physics, 
their systematic applicability to explain real-world collective motion patterns, as well as   
the assumption of the existence of a velocity-alignment mechanism behind all active systems displaying collective effects,  
has been called into question by a series of pioneering works~\cite{sano1996,romanczuk2009,strombom2011,moussaid2011,pearce2014,huepe2013,huepe2015,grossmann2013,soto2014}.  
In particular, it has been recently shown in~\cite{barberis2016} that 
active particles that interact only by a short-range, position-based, attractive force that acts inside a vision cone (VC) display 
various large-scale self-organized patterns: aggregates, nematic bands, and moving, locally polar structures referred to as worms (see Fig.~\ref{fig:patterns}). 
The resemblance of these emerging patterns to some self-organized behaviors found in nature~\cite{moussaid2011,calovi2014,toulet2015}, 
together with the simplicity of the model, making it amenable to analytical treatments, places position-based flocking models as serious candidates to both describe 
real-world active systems and address fundamental theoretical questions of nonequilibrium (active) systems. 
Here, we review and extend the derivation of the hydrodynamic equations first outlined in~\cite{barberis2016}. 
We start by providing a definition of the microscopic model, formulated in terms of a Langevin equation (Sect.~\ref{sec:micro})  
to later search for a coarse-grained description of the model by deriving the corresponding nonlinear Fokker--Planck equation and performing a moment expansion (Sect.~\ref{sec:derivHydro}). 
The most subtle step in the derivation of the hydrodynamic equations is the use of local ans{\"a}tze to close the infinite hierarchy of equations for the obtained fields (Sect.~\ref{sec:closure}). 
The procedure allows us to unveil the three distinct nontrivial macroscopic behaviors of the system: a) aggregate formation in the absence of orientation order, b) the emergence of nematic bands, 
and c) the appearance of locally polar structures called worms (Fig.~\ref{fig:patterns}).  
We find that (a) can be described by only one macroscopic field, the density, while (b) and (c) require at least two fields: density and local nematic order for (b), and density and local polar order for (c). 
The analysis indicates that position-based flocking models are fundamentally different from other active systems, including 
velocity-alignment-based flocking systems~\cite{vicsek1995,toner1995,toner1998,chate2008,toner2012,ramaswamy2003,chate2006,ngo2014,peruani2006,peruani2008,baskaran2008,ginelli2010,peshkov2012b,abkenar2013,weitz2015,nishiguchi2016}.

\section{Microscopic model} \label{sec:micro}

\subsection{Equations of motion}

\begin{figure}
 \begin{center}
 \resizebox{4cm}{!}{\rotatebox{0}{\includegraphics{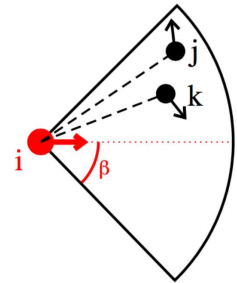}}}
 \caption{(Color online) Sketch illustrating the model defined by Eq.~(\ref{eq:acceleration_full}), whose dynamics in two dimensions reduces 
 to that given by Eq.~(\ref{eq:mov_2d}). 
 Particle positions are indicated by circles and their velocities by arrows. 
 Particles interact with  particles inside the VC. In the figure, the VC of particle $i$ is displayed. 
 Notice that the 
 orientation of the VC is given by particle $i$'s velocity (red arrow). 
 In the sketch, particle $i$ interacts exclusively with particles $j$ and $k$, and    
 only the  positions of $j$ and $k$, and not their velocities,  are relevant for the evolution of $i$.
 The state of particle $i$ is given by its position $\mathbf{x}_{i}$ 
 and its velocity, which is parametrized in two dimensions by only the angle $\theta_i$ since the dynamics keeps the speed  constant.
 For more details on the model, see the text. 
 }
 \label{fig:sketch}
 \end{center}
 \end{figure}

We consider particles moving at a constant speed, which means 
that any acceleration experienced by  a particle occurs in the direction perpendicular to 
its instantaneous velocity. 
Given the constraint imposed on the particles, i.e., moving at constant speed, the 
equation of motion of the $i$th particle in any dimension is given by 
\begin{align}
\label{eq:acceleration_full} 
\ddot{\mathbf{x}}_i = \mathbfcal{P}_i \left( \mathbf{F}_i +  \mathbfcal{N}_i \right) =  -  C_0 \, \dot{\mathbf{x}}_i \times \dot{\mathbf{x}}_i \times \left(\mathbf{F}_i  + \mathbfcal{N}_i \right)   \, ,  
\end{align}
where we have introduced the projector operator $ \mathbfcal{P}_i = -  C_0 \, \dot{\mathbf{x}}_i\! \times\! \dot{\mathbf{x}}_i \times $, with $C_0 = \left[m_i v_0^2\right]^{-1}$ to ensure  that the speed remains constant and equal to $||\dot{\mathbf{x}}_i(t=0)||=v_{0i}$. In Eq.~(\ref{eq:acceleration_full}) $\mathbfcal{N}_i$ denotes  a random force and $\mathbf{F}_i$  an interaction  force. 
Here, we focus on particles that interact via an attractive force that acts inside a vision cone (VC) 
and thus define the force on particle $i$ as:
\begin{align}
\label{eq:force_attract} 
\mathbf{F}_i = \tilde{\gamma}  \sum_{j \in \Omega_i}    \frac{\mathbf{x}_{j} - \mathbf{x}_{i}}{||\mathbf{x}_{j} - \mathbf{x}_{i}||} \, ,
\end{align}
where  $\Omega_i$ denotes the set of neighbors inside the VC of  particle $i$  and $\tilde{\gamma}$ is a constant. 
Particles in the VC are those that satisfy  $||\mathbf{x}_j\! -\! \mathbf{x}_i||\leq R_0$ 
and  $\frac{\mathbf{x}_j - \mathbf{x}_i}{||\mathbf{x}_j - \mathbf{x}_i||} . \left({\dot{\mathbf{x}}_i}/{||\dot{\mathbf{x}}_i||}\right)>\cos(\beta)$, with $\beta$ the {\it size} of the cone. 
This means that, by definition, the cone is oriented in the direction given by $\dot{\mathbf{x}}_i$; for a sketch of the model see Fig.~\ref{fig:sketch}.  
Notice that in Eq.~(\ref{eq:force_attract}) we do not divide by the number of neighbors in contrast to the model analyzed in~\cite{barberis2016}. 
In the following, we assume for simplicity that particles are identical and start with the same speed, such that  $m_i = m_0$ and ${v_0}_i =v_0$ for all $i$. 

\subsection{Dynamics in two dimensions}
 
In order to  simplify the derivation of hydrodynamic equations, in the following we restrict the motion of particles to the two-dimensional plane  $\hat{\mathbf{e}}_1$-$\hat{\mathbf{e}}_2$ 
by assuming that at $t=0$  the velocity of all particles lies on this plane.  
To ensure two-dimensional motion, we additionally require that 
 $\mathbfcal{N}_i(t)$ lies on the plane $\hat{\mathbf{e}}_1$-$\hat{\mathbf{e}}_2$ 
 such that 
$ -  C_0 \,  \dot{\mathbf{x}}_i \times \mathbfcal{N}_i = \sqrt{2 D_{\theta}} \xi_i(t) \hat{\mathbf{e}}_3$, 
with $\langle \xi_i(t) \rangle = 0$ and  $\langle \xi_i(t) \xi_j(t') \rangle = \delta_{i,j} \delta(t-t')$. 
Since we are on a plane and the speed is conserved, we can write $\dot{\mathbf{x}}_i  =  v_0 \mathbf{V}(\theta_i)$ with $\mathbf{V}(.)\equiv (\cos(.),\sin(.))^{T}$  
and thus $\ddot{\mathbf{x}}_i  =  \dot{\theta_i}\,v_0\, \mathbf{V}_{\perp}(\theta_i)$, where $\mathbf{V}_{\perp}(.)\equiv (-\sin(.),\cos(.))^{T}$. 
Using these definitions, Eq.~(\ref{eq:acceleration_full}) can be rewritten as 
\begin{subequations}
\label{eq:mov_2d}
\begin{align}
\label{eq:posvel}
\dot{\mathbf{x}}_i  &=  v_0 \mathbf{V}(\theta_i)\\
\label{eq:theta}
  \dot{\theta}_i &= \gamma \sum_{j \in \Omega_i} T_{ij}+ \sqrt{2D_{\theta}}\xi_i(t)  \, ,
\end{align}  
\end{subequations}
where the angle $\theta_i$ represents the moving direction of the particle on the plane $\hat{\mathbf{e}}_1$-$\hat{\mathbf{e}}_2$    
 and $T_{ij}$ is  defined as  $T_{ij}=\left[\mathbf{V}(\theta_{i})\times \mathbf{V}(\alpha_{ij})\right].\hat{\mathbf{e}}_3 =\sin(\alpha_{ij}-\theta_i)$ with $\mathbf{V}(\alpha_{ij})=\frac{\mathbf{x}_j - \mathbf{x}_i}{||\mathbf{x}_j - \mathbf{x}_i||}$, 
 and $\gamma = \tilde{\gamma}/(v_0 m_0)$.
%

\section{Derivation of hydrodynamic equations} \label{sec:derivHydro}

Since the microscopic model given by Eq.~(\ref{eq:mov_2d}) has been formulated in terms of Langevin equations, 
it is natural to attempt a hydrodynamic description of the system dynamics by deriving 
the corresponding  nonlinear Fokker--Planck equation for $p(\mathbf{x}, \theta, t) = \langle \sum_{i=1}^N \delta(\mathbf{x} - \mathbf{x}_i) \delta(\theta - \theta_i) \rangle$, which reads 
\begin{eqnarray}
\label{eq:FP}
 \partial_t p + \mathbf{\nabla} \left[ v_0 \mathbf{V}(\theta) p\right]   =  D_{\theta} \partial_{\theta \theta} p - \partial_{\theta}\left[ \mathcal{I} p \right]\, ,  
\end{eqnarray}
where $\mathcal{I}$ represents the (average) interaction experienced by a particle located at position $\mathbf{x}$ and 
with moving direction $\theta$ at time $t$. 
The term $\mathcal{I}$ is simply defined as 
%
\begin{eqnarray}
\nonumber
\mathcal{I} &=& \gamma \int_{\Omega(\mathbf{x},\theta)} d\mathbf{x}' \int_0^{2\pi} d\theta' \sin(\alpha(\mathbf{x}'\!\!-\!\mathbf{x}) - \theta) p(\mathbf{x}', \theta', t) \\
\label{eq:interaction_FP} 
&=& \gamma \int_{\Omega(\mathbf{x},\theta)} d\mathbf{x}' \sin(\alpha(\mathbf{x}'\!\!-\!\mathbf{x}) - \theta) \rho(\mathbf{x}',  t)\,
\end{eqnarray}
%
where $\Omega(\mathbf{x},\theta)$ corresponds to the VC for a particle located at $\mathbf{x}$ moving in direction $\theta$,  $\alpha(\mathbf{x}'\!\!-\!\mathbf{x})$ corresponds 
to the angle in polar coordinates of the vector $(\mathbf{x}'\!\!-\!\mathbf{x})/||\mathbf{x}'\!\!-\!\mathbf{x}||=\mathbf{V}(\alpha)$, 
and where we have introduced the definition $\rho(\mathbf{x},t) = \int^{2\pi}_0 d\theta \, p(\mathbf{x},\theta,t)$. 
Notice that in Eq.~(\ref{eq:FP}) we have assumed that 
 $p_2(\mathbf{x}, \theta, \mathbf{x}', \theta', t) \simeq p(\mathbf{x}, \theta,t) p(\mathbf{x}', \theta',t)$.
We can simplify the calculations by explicitly using $\mathbf{x}'\!\! - \!\mathbf{x}=R \mathbf{V}(\alpha)$, which lets us rewrite the integral over  $\Omega(\mathbf{x},\theta)$ as 
%
%
\begin{eqnarray}
\label{eq:interaction_FP_simple} 
\mathcal{I} = \gamma \int_0^{R_0} dR \, \int_{\theta-\beta}^{\theta+\beta} d\alpha\, R \,\sin(\alpha - \theta) \rho(\mathbf{x}+R \mathbf{V}(\alpha), t).\,
\end{eqnarray}
%
Our next step is  to approximate $\rho(\mathbf{x}+R \mathbf{V}(\alpha), t)\simeq \sum_{0<n+k\leq N} \frac{\partial^{n+k}\rho}{\partial_x^n \partial_y^k} R^{n+k} \frac{\cos(\alpha)^n}{n!} \frac{\sin(\alpha)^k}{k!}$ and insert it into Eq.~(\ref{eq:interaction_FP_simple}) to express $\mathcal{I}$ up to order $R^2$ as  
%
\begin{eqnarray}
\label{eq:interaction_FP_gf} 
\mathcal{I} = \gamma [g(\beta,R_0)  \left(-\partial_x\rho \sin(\theta)  + \partial_y \rho \cos(\theta)\right) \\
\nonumber
+ f(\beta, R_0) ( \partial_{xy}\rho \cos(2\theta)  + (\partial_{yy}\rho  - \partial_{xx}\rho) \frac{\sin(2\theta)}{2})]\, ,
\end{eqnarray}
%
where 
$g(\beta) = (R_0^3/3) \left( \beta - \sin(2 \beta)/2 \right)$ and $f(\beta) = (R_0^4/6) \sin^3(\beta)$. 
Our goal now is to obtain a description of the system in terms of fields that depend on $\mathbf{x}$ and $t$, eliminating the dependence on $\theta$. In order to do this, we multiply the left- and right-hand sides of Eq.~(\ref{eq:FP}) by $\mathbf{V}(k\theta)$, with $k \in \mathbb{N}$, after replacing $\mathcal{I}$ with Eq.~(\ref{eq:interaction_FP_gf}), and integrate over $\theta$. For a compact notation, we introduce the following fields:
%
\begin{subequations}
\label{eq:definition}
\begin{align}
\label{eq:def_P}
\mathbf{P}(\mathbf{x},t)&= \left[ 
\begin{array}{c} 
P_x \\ 
P_y 
 \end{array}
\right]
= \int^{2 \pi}_0 d\theta \, \mathbf{V}(\theta) \, p(\mathbf{x},\theta,t) ,\\
\label{eq:def_Q}
 \mathbf{Q}(\mathbf{x},t)  &= 
  \left[ 
\begin{array}{c} 
Q_c \\ 
Q_s 
 \end{array}
\right]
 =
 \int^{2 \pi}_0 d\theta \, \mathbf{V}(2\theta) \, p(\mathbf{x},\theta,t), \\
\label{eq:def_M}
 \mathbf{M}_q(\mathbf{x},t)  &= 
  \left[ 
\begin{array}{c} 
M_{qc} \\ 
M_{qs} 
 \end{array}
\right]
 =
  \int^{2 \pi}_0 d\theta \, \mathbf{V}(q\theta) \, p(\mathbf{x},\theta,t) ,
 \end{align} 
\end{subequations}
%
with $q$  a natural number greater than $2$. The procedure leads to the following temporal evolution of the fields:
\begin{widetext}
\begin{subequations}
\label{eq:fields}
\begin{align}
\label{eq:fields_1}
\partial_t \rho+v_0 \nabla \cdot \mathbf{P}&=0 \\
\label{eq:fields_2}
\partial_t \mathbf{P} + \frac{v_0}{2} \left(\nabla \rho + \left[ \nabla^T \overline{\overline{\mathcal{M}}}_Q\right]^T \right) &= -D_{\theta} \mathbf{P}  
 - \frac{\gamma g(\beta)}{2} \left[  \overline{\overline{\mathcal{M}}}_Q - \rho \mathbb{1}  \right] \nabla \rho 
- \frac{\gamma f(\beta)}{2} \overline{\overline{\mathcal{M}}}_{\rho1} \left[ \mathbf{P} - \mathbf{M}_3 \right] \\
 \label{eq:fields_3}
 \partial_t \mathbf{Q} + \frac{v_0}{2} \left[\mathbf{\nabla}^T \left(
\overline{\overline{\mathcal{M}}}_3 + \overline{\overline{\mathcal{M}}}_P  
 \right)\right]^T &=
 -4 D_{\theta} \mathbf{Q} 
 - \gamma g(\beta) \left[ \overline{\overline{\mathcal{M}}}_3 - \overline{\overline{\mathcal{M}}}_P^{T} \right]  \nabla \rho 
 - \gamma f(\beta) \left( \overline{\overline{\mathcal{M}}}_{\rho2} \mathbf{M}_4 +  \rho 
\left[ 
\begin{array}{c} 
\Phi \rho \\ 
-\partial_{xy}\rho 
 \end{array}
\right] \right) \, ,
  \end{align} 
\end{subequations}
\end{widetext}
where the symbols $\overline{\overline{\mathcal{M}}}_A$ denote matrices defined 
using the auxiliary matrices  $\mathbb{E}_1 = \left[ \begin{array}{cc} 1 & 0 \\ 0 & -1 \end{array} \right]$, 
$\mathbb{E}_2 = \left[ \begin{array}{cc} 0 & 1 \\ 1 & 0 \end{array}  \right]$, $\mathbb{E}_3 = \left[ \begin{array}{cc} 0 & 1 \\ -1 & 0  \end{array} \right]$, and the unity matrix $\mathbb{1}$ as  
$\overline{\overline{\mathcal{M}}}_Q = Q_c \mathbb{E}_1 + Q_s \mathbb{E}_2$, $\overline{\overline{\mathcal{M}}}_3 = M_{3c} \mathbb{E}_1 + M_{3s} \mathbb{E}_2$, 
$\overline{\overline{\mathcal{M}}}_P = P_x  \mathbb{E}_2 +  P_y  \mathbb{E}_3$, $\overline{\overline{\mathcal{M}}}_{\rho1} = \Phi \rho/2  \mathbb{E}_1 - \partial_{xy}\rho   \mathbb{1}$, and 
$\overline{\overline{\mathcal{M}}}_{\rho2} = \partial_{xy} \rho \mathbb{E}_2 - \Phi\rho/2   \mathbb{E}_1$. 
In addition, we have defined $\Phi \rho$ as $\Phi \rho = \partial_{yy} \rho - \partial_{xx} \rho$.

\begin{figure}
 \begin{center}
 \resizebox{8cm}{!}{\rotatebox{0}{\includegraphics{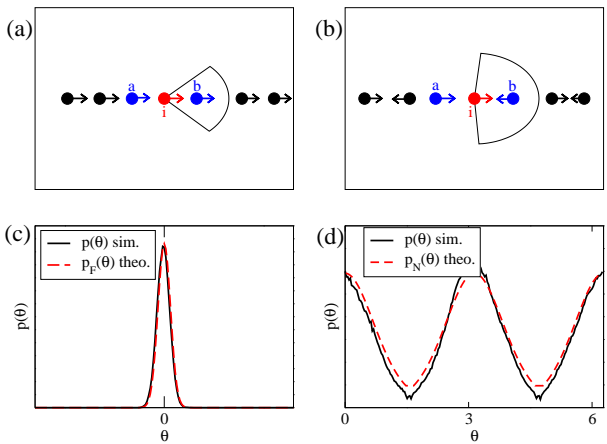}}}
 \caption{(Color online) Emergence of local (orientational) order from a given configuration of particles in space: 
 (a)  Local polar order ($\beta=0.8$, $\sqrt{2 D_{\theta}}=0.18$) and (b)  local nematic order ($\beta=1.5$, $\sqrt{2 D_{\theta}} = 0.75$). 
The positions of particles are indicated by dots and their velocities by arrows. 
Only the VC of particle $i$ is shown. 
Particles $a$ and $b$ are the nearest neighbors, in distance, of particle $i$. By reorienting the VC, particle 
$i$ can interact with either particle $a$ or $b$. 
Notice that the idealized arrangement of particles analyzed here, a straight line of particles, is used only as an illustration. 
Any long-lived, elongated distribution of particles in space will lead to either polar or nematic (local) orientational order.      
(c) [(d)] Angular distribution obtained from simulations of the scenario depicted in (a) [(b)] (solid black curve), which is compared with $p_F(\theta)$ [$p_N(\theta)$] (dashed red curve), Eq.~(\ref{eq:panel_sol}) [Eq.~(\ref{eq:panel_b_nemaSol})].
 }
 \label{fig:local_order}
 \end{center}
 \end{figure}

\section{From local solutions to closure assumptions}\label{sec:closure}

The system of Eq.~(\ref{eq:fields}), owing to the presence of higher-order fields, specifically $\mathbf{M}_3$ and $\mathbf{M}_4$, does not represent a 
closed system of equations. 
If we derive equations for $\partial_t\mathbf{M}_3$ and $\partial_t\mathbf{M}_4$, we will quickly find that they depend on $\mathbf{M}_5$ and $\mathbf{M}_6$. 
In short, we have an infinite hierarchy of equations. In order to work with Eq.~(\ref{eq:fields}), we are forced to find suitable closure assumptions.  
We will make use of local solution ans{\"a}tze to express higher-order fields in terms of $\rho$, $\mathbf{P}$, and $\mathbf{Q}$ and obtain a closed system of equations. 

\subsection{When no local (orientational) order is possible}

We start with a trivial limit. For $\beta = 0$ it is evident that $f=g=0$ and no local orientational order is  possible, i.e.,  $\mathbf{P}=\mathbf{Q}=\mathbf{0}$. 
We are left then with a simple system of non-interacting active particles characterized by a
diffusion coefficient $D_{NAP}=v_0^2/(2 D_{\theta})$. 
Our next step is to study the opposite situation, i.e., $\beta = \pi$, which corresponds to isotropic attractive interactions. 
In this limit, $f=0$ but $g>0$, which implies that particles interact among themselves via 
a standard short-range attractive force. 
Such interactions cannot lead to polar or nematic local orientational order.
%
%
The only relevant field in this scenario is $\rho(\mathbf{x},t)$ and our goal is to find an effective equation for $\partial_t \rho$. 
Given the absence of orientational order, and using the faster relaxation of $\mathbf{Q}$ with respect to $\mathbf{P}$, we ignore Eq.~(\ref{eq:fields_3}) 
by assuming that  $\mathbf{Q}=\partial_t\mathbf{Q}=\mathbf{0}$. 
If we have to obtain a nontrivial dynamics, we cannot simply discard Eq.~(\ref{eq:fields_2}), but assume only that $\partial_t\mathbf{P}=\mathbf{0}$. 
By substituting this assumption into Eq.~(\ref{eq:fields_2}), we find that $\mathbf{P}$ has to satisfy 
\begin{eqnarray}
\label{eq:eq_P_360} 
\frac{v_0}{2} \nabla \rho =  - D_{\theta} \mathbf{P} + \frac{\gamma g}{2} \rho \nabla \rho\, ,
\end{eqnarray}
%
from which we obtain an expression for $\mathbf{P}$ that we insert into Eq.~(\ref{eq:fields_1}) to arrive at  
\begin{eqnarray}
\label{eq:phaseSep}
\partial_t \rho =  - \frac{v_0}{2D_{\theta}} \nabla\left[ - v_0 \nabla \rho + \gamma g \rho \nabla \rho  \right] \, .
\end{eqnarray} 
From Eq.~(\ref{eq:phaseSep}) we learn that a homogeneous spatial distribution of particles becomes  linearly unstable when $c_1 = D_{NAP} - v_0 \gamma g \rho_0/(2 D_{\theta}) <0$. 
This result is obtained by substituting $\rho = \rho_0 + \epsilon \,\delta\!\rho\!\left[\mathbf{x},t\right]$ into Eq.~(\ref{eq:phaseSep}), with $\rho_0$ a constant, $\epsilon \ll1$,  $\delta \! \rho$ the perturbation function, 
and keeping  terms linear in $\epsilon$. 
If we use as a perturbation $\delta \! \rho = e^{\lambda t} e^{i \mathbf{k}.\mathbf{x}}$, we can easily understand that the dispersion relation of the linearized system is not well behaved. 
This problem is fixed by going one order further in the Taylor expansion of Eq.~(\ref{eq:interaction_FP_simple}), which adds  
the term $\gamma \frac{\pi R_0^5}{40} \left[ \cos(\theta) (\! \partial_{xxy}\rho + \partial_{yyy}\rho\!) -\sin(\theta) (\! \partial_{xxx}\rho + \partial_{yyx}\rho\!) \right]$ to Eq.~(\ref{eq:interaction_FP_gf}). 
By incorporating  third order derivatives, it is easy to show that the dispersion relation is of the form $\lambda = - c_1 \mathbf{k}^2 - c_2 \mathbf{k}^4$, where $c_2 = v_0 \gamma \pi R_0^5 \rho_0/80 >0$, which indicates that the dispersion relation (of the linearized system) is qualitatively similar to that of a Cahn-Hilliard equation. 
In summary, for $\beta \sim \pi$  we expect the system to undergo phase separation following standard coarsening for sufficiently large systems.
%
In simulations, deviations from this behavior are expected as long as the characteristic distance between aggregation centers is smaller or comparable to $v_0 D_{\theta}^{-1}$.

\subsection{In the presence of local nematic order}

For $\beta<\pi$ we can conceive the existence of particle configurations leading to some kind of orientational order. 
Logically, only stable configurations are relevant here.  
Given the proposed microscopic equations, there are two relevant particle configurations to be considered: i) an elongated ``band''  with particles moving along it in both directions  
and ii) a line of particles  where all particles move in the same direction, i.e., where particles follow each other.  
These two configurations  emerge spontaneously in simulations of the microscopic model (see Fig.~\ref{fig:patterns}). 

Our first step is to understand that if we fix particles in space on an elongated high-density structure and apply Eq.~(\ref{eq:theta}),  we  
obtain an asymptotic local distribution $p(\theta)$ displaying nematic symmetry. 
It is important to stress that the idealized configurations shown in Fig.~\ref{fig:local_order} 
serve as an illustration of a generic mechanism leading to orientation order. 
The arguments put forward below hold true for any long-lived  spatial distribution  
of particles that displays high accumulation of particles along a given direction, 
and where each particle may interact with multiple particles simultaneously. 
Thus, for simplicity and without loss of generality we  focus on the idealized situation depicted in Fig.~\ref{fig:local_order}(b).   
Let us start by simulating the dynamics of $\dot{\theta_i}$ as given by Eq.~(\ref{eq:theta}). 
In this configuration, particle $i$ interacts for some time with particle $a$, some time with particle $b$, and some time with neither of them, depending on the orientation of its VC.   
The dynamics of $\theta_i$ is then given by 
\begin{eqnarray}
\label{eq:panel_b} 
\dot{\theta}_i &=& \gamma \sum_{j=\{a,b\}} \sin(\alpha_{i\,j}-\theta_i) h_j(\theta)+ \sqrt{2D_{\theta}} \xi_i(t) \, ,
\end{eqnarray}
where $\alpha_{i\,a}=\pi$ denotes the polar angle of the vector $(\mathbf{x}_a - \mathbf{x}_i)/||\mathbf{x}_a - \mathbf{x}_i||=\mathbf{V}(\alpha_{i\,a})$. Similarly,  
$\alpha_{i\,b}=0$ is associated with the vector $(\mathbf{x}_b - \mathbf{x}_i)/||\mathbf{x}_b - \mathbf{x}_i||=\mathbf{V}(\alpha_{i\,b})$, 
and the VC of particle $i$ is described via  the two auxiliary functions $h_a(\theta_i)$ and  $h_b(\theta_i)$, 
which are defined in such a way that $h_a(\theta_i)=1$ when particle $a$ lies within the VC of $i$, and $0$ otherwise, while  $h_b(\theta_i)=1$ when particle $b$ is located inside the VC of $i$, and $0$ otherwise. 
It is easy to verify that the asymptotic distribution of $\theta_i$ can be 
approximated by $p_{a1}(\theta) \sim \mathcal{N} \left[ \sum_{j} e^{\frac{\gamma}{D_{\theta}} \cos(\alpha_{a\,j}-\theta)} h_j(\theta) + e^{\frac{\gamma}{D_{\theta}}\cos(\beta)} \Pi_{j}(1-h_j(\theta))\right]$, 
with $\mathcal{N}$ a normalization constant.
For simplicity, in the following we focus on large values of $\beta$ and $D_{\theta}$ and approximate 
the dynamics of $\theta_i$ by 
\begin{eqnarray}
\label{eq:panel_b_nema_evol} 
\dot{\theta}_i &=& \gamma \sin\left(2(\alpha-\theta_i)\right) + \sqrt{2D_{\theta}} \xi_i(t) \, ,
\end{eqnarray}
 with $\alpha$ either $0$ or $\pi$ to apply the equation to the configuration sketched in Fig.~\ref{fig:local_order}(b). 
 The advantage of Eq.~(\ref{eq:panel_b_nema_evol}) is that we ignore the difficulties associated with the VC. 
Its associated Fokker--Planck equation reads 
\begin{eqnarray}
\label{eq:panel_b_nema} 
\partial_t p = \gamma \partial_{\theta}\left[ \sin\left(2(\alpha-\theta)\right) p\right]+ D_{\theta} \partial_{\theta\theta} p \, ,
\end{eqnarray}
whose steady-state solution  is the von Mises distribution  
\begin{eqnarray}
\label{eq:panel_b_nemaSol} 
p_{N}(\theta) = \mathcal{N} e^{\frac{\gamma}{2 D_{\theta}} \cos\left( 2(\alpha -\theta ) \right)} =  \mathcal{N} e^{\frac{\gamma}{2 D_{\theta}} \mathbf{V}(2\theta)\cdot \mathbf{V}(2\alpha) }   \, ,
\end{eqnarray}
where $\mathcal{N}$ is again a normalization constant. 
It is evident that $p_{a1}(\theta)$ and $p_{N}(\theta)$ share the same symmetry. Given the many approximations performed to arrive at $p_{N}(\theta)$, it is far from evident that 
Eq.~(\ref{eq:panel_b_nemaSol}) provides a reasonable description of the dynamics defined by Eq.~(\ref{eq:panel_b}). 
Figure~\ref{fig:local_order}(b) shows that $p_N(\theta)$ is a reasonable approximation of the distribution 
 $p(\theta)$ obtained from direct simulations using Eq.~(\ref{eq:panel_b}). 

From the previous arguments we have learned that if particles are arranged in an elongated, high-density spatial configuration, we can expect local nematic order $\mathbf{Q}$ to emerge.  
Notice that Eq.~(\ref{eq:panel_b_nemaSol}) allows us to establish that $\mathbf{Q} \propto \mathbf{V}(2\alpha)$. 
We  use this knowledge to conceive the closure of the derived field equations, i.e., Eq.~(\ref{eq:fields}).  
Assuming that the dynamics of $\theta$ is faster than the spatial dynamics, we expect that locally the distribution of $\theta$ will follow the functional form suggested by Eq.~(\ref{eq:panel_b_nemaSol}), which we write 
generically as  
\begin{eqnarray}
\label{eq:dist_local_nema} 
p(\mathbf{x}, \theta,t) =  \mathcal{N} e^{w \mathbf{V}(2\theta)\cdot \mathbf{Q} }   \, ,
\end{eqnarray}
where $\mathcal{N}$ as well as $w$ may  depend on  $\rho$. 
Notice that to simplify the notation we have not written 
 the dependence of $\rho$ and $\mathbf{Q}$ on $\mathbf{x}$ and $t$.  
Expressions for $\mathcal{N}$ and $w$ can be obtained by self-consistency since, by definition, $p(\mathbf{x}, \theta,t)$ has to obey 
\begin{eqnarray}
\label{eq:density} 
\int_0^{2\pi} d\theta\,  p(\mathbf{x}, \theta,t) \sim \mathcal{N} 2\pi= \rho  \, , 
\end{eqnarray}
and thus $\mathcal{N} = \frac{\rho}{2\pi}$, while from the definition of $\mathbf{Q}$ we find 
\begin{eqnarray}
\label{eq:nema} 
\int_0^{2\pi} \!\!d\theta\, \mathbf{V}(2\theta) p(\mathbf{x}, \theta,t) &=&\\
\nonumber
 &&\!\! \int_0^{2\pi} \!\!d\theta\, \mathbf{V}(2\theta) \mathcal{N} e^{w\mathbf{V}(2\theta)\cdot\mathbf{Q}} \sim \mathbf{Q}  \, , 
%
\end{eqnarray}
which leads to $w=\frac{2}{\rho}$. 
All this means that our local ansatz reads 
\begin{eqnarray}
\label{eq:nema_ansatz} 
p(\mathbf{x},\theta,t) = \frac{\rho(\mathbf{x},t)}{2\pi} e^{\frac{2}{\rho(\mathbf{x},t)} \mathbf{Q}(\mathbf{x},t)\cdot\mathbf{V}(2\theta)}\,.  
\end{eqnarray} 
This approximation is valid close to the onset of local order, i.e., when $||\mathbf{Q}/\rho||$ is small. 
With $p(\mathbf{x},\theta,t)$ at hand, we can compute all the remaining fields: $\mathbf{M}_3$ and $\mathbf{M}_4$.  
By symmetry, it is easy to verify that $\mathbf{M}_3(\mathbf{x},t)\sim \mathbf{0}$.  
Strictly speaking, we can show that  $\mathbf{M}_3$ is of order higher than $\mathcal{O}(\mathbf{Q}^4)$.  
The only remaining field to analyze is $\mathbf{M}_4(\mathbf{x},t)$.  
By subsituting the local ansatz into its definition, we find 
%
\begin{eqnarray}
\nonumber
\mathbf{M}_4(\mathbf{x},t) &=&   \int_{0}^{2\pi} d\theta \mathbf{V}(4\theta)  \frac{\rho(\mathbf{x},t)}{2\pi} e^{\frac{2}{\rho(\mathbf{x},t)} \mathbf{Q}(\mathbf{x},t)\cdot\mathbf{V}(2\theta)} \\
\label{eq:consistency_nematic}
&=& \frac{1}{\rho(\mathbf{x},t)}   \left[ \begin{array}{c} \frac{1}{2}(Q_c(\mathbf{x},t)^2 - Q_s(\mathbf{x},t)^2) \\ Q_c(\mathbf{x},t) Q_s(\mathbf{x},t) \end{array}  \right]\! .
\end{eqnarray}
%
By neglecting $\mathbf{M}_3$ and using Eq.~(\ref{eq:consistency_nematic}), Eq.~(\ref{eq:fields}) becomes a closed system. 
Furthermore, the system dynamics can be reduced to the evolution of only two fields:  $\rho$ and $\mathbf{Q}$. 
In order to do this, we require $\partial_t \mathbf{P}=0$ at all times, which allows the fusing of Eqs.~(\ref{eq:fields_1}) and~(\ref{eq:fields_2}),  
and keep the leading-order terms in Eq.~(\ref{eq:fields_3}). This  procedure leads to:
%
\begin{subequations}
\label{eq:fields_simplified}
\begin{align}
&\partial_t \rho =   \\
\nonumber
&v_0 \nabla \cdot \left[ -\overline{\overline{\mathcal{M}}}_{I} \left(  \frac{v_0}{2} \left(\nabla \rho + \left[ \nabla^T \overline{\overline{\mathcal{M}}}_Q\right]^T \right) + \frac{\gamma g}{2} \left[  \overline{\overline{\mathcal{M}}}_Q - \rho \mathbb{1} \right] \nabla \rho  \right) \right]\\
 &\partial_t \mathbf{Q} +4 D_{\theta} \mathbf{Q}  = 
 \\
 \nonumber
  & - \gamma f \left( \overline{\overline{\mathcal{M}}}_{\rho2} 
\frac{1}{\rho} \left[ 
\begin{array}{c} 
\frac{1}{2} \left(Q_c^2 - Q_s^2 \right) \\ 
Q_c Q_s 
 \end{array}
 \right] 
  +   
\rho \left[ 
\begin{array}{c} 
\Phi \rho \\ 
-\partial_{xy}\rho 
 \end{array}
\right] \right) \, ,
\end{align} 
\end{subequations}
where $\overline{\overline{\mathcal{M}}}_{I}$ is the inverse of $D_{\theta}\mathbb{1} + \frac{\gamma f}{2}\overline{\overline{\mathcal{M}}}_{\rho1}$.

Since our goal is to look for static self-organized nematic patterns, we do not need to consider the temporal evolution of the fields. Moreover, we search for steady-state solutions and thus set all partial temporal derivatives equal to zero. We apply this condition to Eq.~(\ref{eq:fields}). 
Given that all directions are equivalent, without loss of generality we assume that nematic order occurs along the $\hat{x}$-axis,  i.e., $Q_s=0$. 
This implies that the pattern is invariant along the $\hat{x}$-axis, an assumption consistent with the nematic bands found in agent-based simulations [Fig.~\ref{fig:patterns}(b)].  
As a consequence of such  invariance, all derivatives with respect to $x$ vanish and fields cannot depend on $x$, 
which, together with the assumption of nematic order along the $\hat{x}$-axis, yields  $p(\mathbf{x},\theta) = \frac{\rho(y)}{2\pi} e^{2Q_c(y)\cos(2\theta)/\rho(y)}$.  
We have already pointed out that the presence of local nematic order implies that $\mathbf{M}_3 \sim \mathbf{0}$,  
and from Eq.~(\ref{eq:consistency_nematic}) 
we learn that  $M_{4c} = \frac{Q_c(y)^2}{2\rho(y)}$ and $M_{4s}=0$. 
Under these assumptions, it is easy to verify that Eq.~(\ref{eq:fields_1}) is automatically satisfied as it occurs for the equations for $P_x$ and $Q_s$ [see Eqs.~(\ref{eq:fields_2}) and ~(\ref{eq:fields_3}), respectively]. 
We are left with the equation for $P_y$ and $Q_c$, which reads 
%
\begin{subequations}
\label{eq:bands}
\begin{align}
\label{eq:bands_py}
\frac{v_0}{2} \partial_y\left[ \rho - Q_c\right] = \gamma \,g(\beta) \partial_y \rho \, \frac{\rho+Q_c}{2} \\
\label{eq:bands_qc}
0 = - 4 D_{\theta} Q_c - \frac{\gamma}{2} \, f(\beta) \partial_{yy} \rho \left( \rho - \frac{Q_c^2}{2\rho} \right) \,.
\end{align} 
\end{subequations}
%
These equations can be expressed as the following first-order ordinary differential equation (ODE) system:
%
\begin{subequations}
\label{eq:firstOrder}
\begin{align}
\label{eq:firstOrder1}
\partial_y z &= -Q_c \left[b \left(\rho - \frac{Q_c^2}{2\rho} \right)\right]^{-1} \\
\label{eq:firstOrder2}
\partial_y \rho &= z \\
\label{eq:firstOrder3}
\partial_y Q_c &= \left( 1 - a (\rho+Q_c) \right) z  \, ,
\end{align} 
\end{subequations}
%
where we have introduced the auxiliary field $z$, given by Eq.~(\ref{eq:firstOrder2}), and the constants $a=\frac{\gamma g}{v_0}$ and $b=\frac{\gamma f}{8D_{\theta}}$. 
Our next step is to linearize either Eq.~(\ref{eq:bands}) or Eq.~(\ref{eq:firstOrder}) using $\rho(y) = \rho_0 + \epsilon \delta \rho(y)$ and  $Q_c =  \epsilon \delta Q_c(y)$, 
with $\rho_0$ a constant that represents a linear density and $\epsilon$ a perturbation parameter such that $\epsilon \ll 1$, with $\delta\! \rho(y)$ and $\delta Q_c(y)$ 
the perturbation functions to be determined. Keeping the linear-order terms with respect to $\epsilon$, it is possible to show that the linear system reduces to 
\begin{equation}
\label{eq:oscillator} 
\partial_{yy} \tilde{z} = \frac{a \rho_0 - 1}{b \rho_0} \tilde{z} \, ,
\end{equation}
where $\tilde{z}=\partial_{y}\delta \rho$. This reduction is possible because $\delta Q_c = - b \rho_0 \partial_{yy} \delta \rho$ in the linearized system. 
It is evident that Eq.~(\ref{eq:oscillator}) admits  trigonometric functions as solutions  when $a \rho_0 - 1 <0$. This implies that we expect the  presence of multiple nematic parallel bands, where density and nematic order are closely related: $Q_c(y) \propto \rho(y)$.

In summary, by assuming the presence of local nematic order, we obtained a closed system of field equations  
and showed that this system of equations has steady-state solutions. Furthermore, we  indicated that these solutions are consistent with the presence of multiple parallel nematic bands  observed in agent-based simulations. 

\subsection{In the presence of local polar order}

Particles arranged in elongated spatial  configurations can also exhibit (transient~\cite{note1}) local polar order. 
In agent-based simulations it becomes evident that elongated particle configurations with polar order exhibit long-lived,  dynamical structures that we refer to as worms. 
Our first goal is to understand how an elongated configuration of particles can induce a local distribution of $\theta$ displaying polar symmetry. 
We start by looking at the configuration shown in Fig.~\ref{fig:local_order}(a). We stress that the idealized configuration depicted in 
Fig.~\ref{fig:local_order}(a) only serves as an illustration of a generic orientational order mechanism. 
To further simplify the argument we ignore  particle $a$ and express the dynamics of $\theta_i$ as 
\begin{eqnarray}
\label{eq:panel_a} 
\dot{\theta}_i &=& \gamma\sin(\alpha_{i\,b}-\theta_i) h_b(\theta) + \sqrt{2D_{\theta}} \xi_i(t) \, .
\end{eqnarray}
%
%
%
The associated Fokker--Planck equation of Eq.~(\ref{eq:panel_a}) -- for $h_b(\theta)=1$ for all $\theta$ -- reads:
\begin{eqnarray}
\label{eq:panel_aFP} 
\partial_{t}p(\theta,t) &=& - \partial_{\theta}\left[ \gamma \sin(\alpha_{i\,b} - \theta) p - D_{\theta} \partial_{\theta} p \right] \, .
\end{eqnarray}
The steady-state solution of Eq.~(\ref{eq:panel_aFP}), denoted by $p(\theta)$, takes the form 
\begin{eqnarray}
\label{eq:panel_sol} 
p_F(\theta) &=& \mathcal{N} e^{\frac{\gamma}{D_{\theta}}\cos(\theta)} =  \mathcal{N} e^{\frac{\gamma}{D_{\theta}}\mathbf{V}(\theta)\cdot\mathbf{V}(\alpha_{i\,b})}\, ,
\end{eqnarray}
where $\mathcal{N}$ is again a normalization constant. Now, we observe that in the idealized image depicted in Fig.~\ref{fig:local_order}(a), the nearest neighbors  of $i$ share the same orientation (see arrows) and so the local polar order $\mathbf{P}$ is parallel to $\mathbf{V}(\alpha_{i\,b})$. 
%
%
The previous assumption allows us to express  the solution given by Eq.~(\ref{eq:panel_sol}) as 
$p(\theta,t)=\mathcal{N} e^{w\mathbf{V}(\theta)\cdot\mathbf{P}}$; see comment below on the estimation of $\mathcal{N}$ and $w$.   
If we analyze the problem with the original definition of $h_b(\theta)$, we find that $p(\theta,t)\simeq\mathcal{N} \left[e^{\frac{\gamma}{D_{\theta}}\mathbf{V}(\theta)\cdot\mathbf{P}} h_b(\theta)+e^{\frac{\gamma}{D_{\theta}}\mathbf{V}(\beta)\cdot\mathbf{P}}(1-h_b(\theta)) \right]$. 
Figure~\ref{fig:local_order}(a) shows that Eq.~(\ref{eq:panel_sol}) is a good approximation of this expression. 
Thus, we adopt the functional form given by Eq.~(\ref{eq:panel_sol}) 
as the local ansatz 
for the distribution of $\theta$.  
After fixing $\mathcal{N}$ and $w$ by requiring $\int d\theta p(\theta,t) = \rho(\mathbf{x}, \theta, t)$ and $\int d\theta \mathbf{V}(\theta)\,p(\theta,t) = \mathbf{P}(\mathbf{x}, \theta, t)$, we find 
\begin{eqnarray}
\label{eq:local_polar} 
p(\mathbf{x}, \theta, t) &=& \frac{\rho(\mathbf{x},t)}{2\pi} e^{\frac{2}{\rho(\mathbf{x},t)} \mathbf{P}(\mathbf{x},t)\cdot\mathbf{V}(\theta)}\, .
\end{eqnarray}

It is important to understand that in this argument we have not considered the motion of particles. We know that for a static, elongated spatial configuration of particles, polar order can only be observed during a transient. However, here the spatial configuration of particles is also evolving. 
In particular, the temporal evolution of the spatial configuration of particles may be such that polar order is maintained. 
Thus, adopting  as the local ansatz Eq.~(\ref{eq:local_polar}), we   compute $\mathbf{Q}$ and $\mathbf{M}_3$ as 
%
\begin{subequations}
\label{eq:consistency_polar}
\begin{align}
\label{eq:consistency_polar_Q}
\mathbf{Q}(\mathbf{x},t) &=   \frac{1}{\rho}   \left[ \begin{array}{c} \frac{1}{2} (P_x^2 - P_y^2)  \\ P_y P_x   \end{array} \right] \\
\label{eq:consistency_polar_m3}
\mathbf{M}_3(\mathbf{x},t) &=   \frac{1}{6\rho^2}   \left[ \begin{array}{c} P_x^3 - 3 P_x P_y^2 \\ 3 P_y P_x^2 - \,\, P_y^3  \end{array}  \right] \, ,
\end{align} 
\end{subequations}
%
where we have not explicitly written the dependence on $\mathbf{x}$ and $t$ for the field $\rho$, $P_x$, and $P_y$. 
Equation~(\ref{eq:consistency_polar}), together with  Eq.~(\ref{eq:fields}), allows us to obtain a closed system of equations for $\rho$ and $\mathbf{P}$,  
%
%
where we have to use the definition of $\mathbf{M}_3$ given above, the definition of $\overline{\overline{\mathcal{M}}}_{\rho1}$ provided 
below Eq.~(\ref{eq:fields}), and the following definition of $\overline{\overline{\mathcal{M}}}_Q$:
\begin{subequations}
\label{eq:newM}
\begin{align}
\nonumber
\overline{\overline{\mathcal{M}}}_Q = \frac{1}{\rho} \left[ \begin{array}{cc} \frac{1}{2} (P_x^2 - P_y^2) & P_x P_y \\ P_x P_y & -\frac{1}{2} (P_x^2 - P_y^2) \end{array} \right]\, .
\end{align} 
\end{subequations}
Let us now investigate the possibility of having static, straight, percolating polar bands. 
Since all directions should be equivalent, for simplicity and without loss of generality,   
we assume that the polar order is along the $x$-axis and the pattern is invariant along $x$. 
This implies that all derivatives with respect to $x$ and time vanish; the latter is due to the fact that we look for static patterns. 
All this together means that 
$P_y=0$, $P_x=P_x(y)$, and $\rho=\rho(y)$.  By inserting this into Eq.~(\ref{eq:fields}), we find that 
Eq.~(\ref{eq:fields_1}) is automatically satisfied, while from Eq.~(\ref{eq:fields_2}) we obtain 
\begin{subequations}
\label{eq:forWorm}
\begin{align}
\label{eq:forWorm_1}
0 &= -D_{\theta} P_x - \frac{\gamma f}{4} \left[ 1 - \frac{1}{6\rho^2} P_x^2 \right] \partial_{yy} \rho \, P_x \\
\label{eq:forWorm_2}
\partial_y \rho &- \frac{1}{2} \partial_y\left( \frac{P_x^2}{\rho}\right) = \frac{\gamma g}{v_0} \partial_y \rho \left( \frac{1}{2}\frac{P_x^2}{\rho} - \rho \right) \,.
\end{align} 
\end{subequations}
From Eq.~(\ref{eq:forWorm_1}) we express $\frac{P_x^2}{\rho}$ as a function of $\rho$ and derivatives of $\rho$. 
The next step is to insert the resulting expression into  Eq.~(\ref{eq:forWorm_2}). 
After performing an expansion in  $\rho$,  we find that Eq.~(\ref{eq:forWorm}) has no solution. 
This proves that percolating, static polar patterns are not a solution of Eq.~(\ref{eq:fields}). 
We stress that this result does not preclude the existence of dynamic locally polar structures such as the dynamic worms observed 
in agent-based simulations.   
%
%
%
In summary, from the hydrodynamic equations we learn that while dynamical, locally polar structures can exist, percolating, static polar bands leading to global polar order
-- i.e., structures similar to the obtained nematic bands discussed above but polar --  cannot emerge.

\section{Conclusions} \label{sec:conclusions}
The derived hydrodynamic equations reveal that 
 active particles interacting only by short-range, position-based, attractive interactions 
can exhibit various complex collective motion patterns if Newton's third law is broken 
by using a vision cone. 
For isotropic, and thus reciprocal interactions, i.e., $\beta=\pi$, we have shown that for sufficiently small $D_{\theta}$ values, the system undergoes  
 phase separation with a classical, equilibrium-like, coarsening dynamics. It is during this phase that we observe the formation of aggregates with no orientation order. 
 This  behavior  is expected to be representative of what happens in the vicinity of isotropic interactions. 
%
For $\beta<\pi$, interactions are nonreciprocal, and for small  $D_{\theta}$ and $\beta$ values, the absence of Newton's third law leads to interesting effects.  
In particular, we have seen that the accumulation of particles along a given direction in space ( i.e., the formation of a high-density stripe), can 
induce either polar or nematic local orientational order.   
We made use of the proposed local distributions of orientations for the polar and nematic cases to obtain suitable closures of the derived hydrodynamic equations. 
We learned that for locally polar structures --  worms -- there is no static, percolating polar band, which indicates the absence of  global polar order. 
For locally nematic structures, on the other hand, we managed to find static patterns, which correspond to elongated  structures leading to global nematic order:  nematic bands. 
All these observations are (qualitatively) consistent with observations with agent-based simulations. 

One important message from the derived hydrodynamic equations is that position-based flocking models belong to a distinct active class and 
are fundamentally different from velocity-alignment-based flocking models, including 
the so-called polar fluids~\cite{toner1995,toner1998,chate2008,bertin2009}, active nematics~\cite{ramaswamy2003,chate2006,ngo2014}, and self-propelled rods~\cite{peruani2006,peruani2008,baskaran2008,ginelli2010,peshkov2012,abkenar2013,weitz2015,nishiguchi2016}. 
For instance, in position-based flocking models such as the one analyzed here, the orientational order that emerges is always  associated with density instabilities,   
and polar or nematic spatially homogeneous ordered phases cannot exist. This is in sharp contrast with the spatially homogeneous Toner--Tu polar phase in 
(velocity-alignment-based)  polar fluids~\cite{toner1995,toner1998,chate2008,bertin2009} and the spatially homogeneous nematic phases reported in~\cite{ramaswamy2003,peruani2008,baskaran2008,ginelli2010,peshkov2012}. 
%


Several fundamental questions remain open for position-based flocking models. The nature of the transitions between the different macroscopic phases has not been explored, 
the effect of macroscopic fluctuations  has not been addressed, a systematic study of the characteristic width of the emerging patterns (worms and nematic bands) is missing, and the impact of boundary conditions has not been analyzed, to name a few of the relevant issues to be clarified. 
In short, very little is known about position-based flocking models, despite the fact that navigation strategies based on positional information may prove key to understanding several biological collective motion patterns~\cite{sano1996,romanczuk2009,moussaid2011,pearce2014,huepe2013,ginelli2015,toulet2015} and in the design of flocking robots~\cite{volpe2015}. 
We expect these active systems to receive considerable attention in the near future. \\

\begin{acknowledgments}

Financial support from Agence Nationale de la Recherche via Grant ANR-15-CE30-0002-01 is acknowledged.

\end{acknowledgments}

\bibliographystyle{apsrev}

\begin{thebibliography}{46}
\bibitem{note1} Strictly speaking, a static, elongated spatial configuration of 
 particles displays asymptotically local nematic order. However, if the motion of particles is taken into account, the dynamics of the spatial configuration can be such 
that polar order is maintained, as observed in agent-based simulations.
\end{thebibliography}

\end{document}